\documentclass[11pt]{article}
\usepackage{latexsym}  
\usepackage{amssymb}
\usepackage{graphicx}
\usepackage{amsmath}

\begin{document}

\begin{center}
{\Large\bf Spectroscopy of Family Gauge Bosons }

\vspace{4mm}
{\bf Yoshio Koide}

{\it Department of Physics, Osaka University, 
Toyonaka, Osaka 560-0043, Japan} \\
{\it E-mail address: koide@kuno-g.phys.sci.osaka-u.ac.jp}

\end{center}

\vspace{3mm}

\begin{abstract}
Spectroscopy of family gauge bosons is investigated 
based on a U(3) family gauge boson model proposed by
Sumino.  In his model, the family gauge bosons are in mass
eigenstates in a diagonal basis of the charged 
lepton mass matrix.  Therefore, the family numbers are defined
by $(e_1,e_2, e_3)=(e, \mu, \tau)$,   
while the assignment for quark sector are free.  
For possible family-number assignments $(q_1, q_2, q_3)$, 
under a constraint from $K^0$-$\bar{K}^0$ mixing, we investigate 
possibilities of new physics, e.g. production of the lightest family gauge boson 
at the LHC, $\mu^- N \rightarrow e^- N$, rare $K$ and $B$ decays, and so on. 
\end{abstract}

PCAC numbers:  
  11.30.Hv, 
  12.60.-i, 
  14.70.Pw, 

\vspace{5mm}

\noindent{\large\bf 1 \ Introduction} 

The most exciting subject in particle physics is to understand 
the origin of ``flavor".
It seems to be very attractive to understand ``families" 
(``generations") in quarks and leptons from concept of 
a symmetry \cite{f_symmetry}. 
Since the observed masses of quarks and leptons are in range 
of $10^{-3}-10^2$ GeV, we may suppose a possibility that 
the lightest family gauge boson can be observed by
terrestrial experiments, e.g. at the LHC. 
   
However, when we try to consider such a visible family gauge 
boson model, we always meet with constraints from the observed 
pseudo-scalar-anti-pseudo-scalar meson mixings $P^0$-$\bar{P}^0$  
($P=K, D, B, B_s$).
The constraints are too tight to allow family gauge bosons with 
lower masses. 
It is usually taken that a scale of 
the symmetry braking is considerably high (e.g. an order of, 
at least, $10^4$ TeV). 
However, there is a family gauge boson model \cite{K-Y_PLB12} in which 
such severe constraints from the $P^0$-$\bar{P}^0$ mixings can be 
considerably loosen.  
In the model, the family gauge symmetry is U(3), 
so that a number of 
the family gauge bosons are nine (not eight), and 
quarks and leptons interacts with the family 
gauge bosons $A_i^{\ j}$ is given by 
 $$
{\cal H}_{fam} = \frac{g_{F}}{\sqrt{2}} \left[ (\bar{e}_i \gamma_\mu e_j) 
+ (\bar{\nu}_i \gamma_\mu \nu_j) 
+ U^{* u}_{ik} U_{jl}^u (\bar{u}_k \gamma_\mu u_l)
+  U^{* d}_{ik} U_{jl}^d(\bar{d}_k \gamma_\mu d_l) 
 \right] (A_i^{\ j})^\mu ,
\eqno(1.1)
$$
where $(u^0_i, d^0_i)$ 
are eigenstates of the family symmetry U(3) and those are
define by $(u^0_i, d^0_i)= (U^u_{ij} u_j, U^d_{ij} d_j)$. 
(The expression (1.1) is based on an extended version \cite{K-Y_PLB12}
of the Sumino model \cite{Sumino_PLB09}. 
See in the next section.)
Note that in the limit of no quark mixing, the family number
is exactly conserved, so that the whole $P^0$-$\bar{P}^0$ mixings are
forbidden. 
(A brief review is given in the next section.)

Another remarkable point in the Sumino model is that 
the family gauge coupling constant $g_F$ and ratios among 
the family gauge boson masses $M_{ij}$ are not free, and 
when once a model is settled, $g_F$ and $M_{ij}/M_{kl}$ 
are fixed. 
Therefore, the model can give a clear answer to observations. 

The family number in the Sumino model \cite{Sumino_PLB09} is 
defined by the charged lepton sector $e_i=(e, \mu, \tau)$ 
and the gauge boson masses 
are given proportionally to the charged lepton masses. 
On the hand, family number in the quark sector may be 
$d^0_i=(d^0, s^0, b^0)$, but it may be an inverted assignment 
$d^0_i=(b^0, s^0, d^0)$, and also a twisted assignment 
$d^0_i=(b^0, d^0, s^0)$. 
(Of course, we consider the same assignments for $u^0_i$ 
because of SU(2)$_L$ symmetry.)
There are six possible assignments of $(u^0_i, d^0_i)$ correspondingly
to  $e_i=(e, \mu, \tau)$.  
(Hereafter, for convenient,  we will denote $q^0_i$ as $q_i$
simply.)

In the present paper, based on the Sumino model \cite{Sumino_PLB09} 
(and also an extended Sumino model \cite{K-Y_PLB12}), 
we investigate visible effects of the family gauge bosons, 
i.e. the  deviations from the $e$-$\mu$-$\tau$ universality, 
rare $K$ and $B$ decays, $\mu$-$e$ conversion, direct production of 
the lightest family gauge boson, and so on.  
We will conclude that the case with a twisted assignment 
$d^0_i=(b^0, d^0, s^0)$ can give rich phenomenology to us.

\vspace{5mm}

\noindent{\large\bf 2 \ Sumino mechanism}

Priori to our investigation, let us give a brief review of 
the Sumino model and its extended version. 

The necessity of the family gauge bosons was first pointed out
by Sumino \cite{Sumino_PLB09}.  
Sumino has paid 
why the charged lepton mass relation \cite{K-mass}
$$ 
K \equiv \frac{m_e + m_\mu +m_\tau}{\left(\sqrt{m_e}+ \sqrt{m_\tau} 
+\sqrt{m_\tau}\right)^2}  = \frac{2}{3} , 
\eqno(2.1)
$$  
is well satisfied by the pole masses (not by the running masses). 
The running masses $m_{ei}(\mu)$ are given by \cite{Arason92}
$$
m_{ei}(\mu) = m_{ei} \left[ 1-\frac{\alpha_{em}(\mu)}{\pi} 
\left( 1 +\frac{3}{4} \log \frac{\mu^2}{m_{ei}^2(\mu)} \right) \right] .
\eqno(2.2)
$$
If the factor $\log(m_{ei}^2/\mu^2)$ in Eq.(2.2) is absent, then the 
running masses $m_{ei}(\mu)$ are also satisfy the formula (2.1).
Sumino has required that contribution of family gauge bosons to the 
charged lepton mass $m_{ei}(\mu)$ cancels the factor $\log(m_{ei}^2/\mu^2)$
due to photon.  
That is, in the collection factors, 
$$
\varepsilon_0 +\varepsilon_i \equiv
e^2 \log \frac{m_{ei}^2}{\mu^2} - 2 \left( \frac{g_F}{\sqrt2}\right)^2 
\log\frac{M_{ii}^2}{\mu^2} ,
\eqno(2.3)
$$
the factor $\varepsilon_i$ must be $\varepsilon_i=0$. 
($\varepsilon_0$ denotes a family-number independent part.)
In the Sumino model, the family gauge boson masses $M_{ii}$ are given
$M_{ii}^2 \propto m_{ei}$, so that we can give $\varepsilon_i =0$
by adjusting the gauge coupling constant $g_F$ suitably.
(The details are given later.)

Note that in the Sumino model, the minus sign for the cancellation
comes form a U(3) assignment of the left-handed and right-handed charged leptons 
$e_L$ and $e_R$, $(e_L, e_R)=({\bf 3}, {\bf 3}^*)$ of U(3). 
As a result, we obtain a somewhat unfamiliar gauge current-current interaction
form
$$
{\cal H}_{fam}^{Sumino} = \frac{g_F}{\sqrt{2}}\sum_{f=u, d, \nu, e} 
\left( \bar{f}_L^i \gamma_\mu f_{Lj} - \bar{f}_{Rj} \gamma_\mu f_R^i
\right) (A_i^{\ j})^\mu .
\eqno(2.4)
$$
However, when the assignment  $(e_L, e_R)=({\bf 3}, {\bf 3}^*)$ is
extended to all quarks and leptons $(f_L, f_R)$, we have unwelcome 
situation:   
(i) The model cannot be anomaly free.
(ii) Effective current-current
interactions with $\Delta N_{fam}=2$ ($N_{fam}$ is a family 
number) appear inevitably.

In order to evade these problems, an extended version of 
the Sumino model (K-Y model) \cite{K-Y_PLB12} 
 has been proposed by Yamashita and the author:
(i) U(3) assignment is $(f_L, f_R)=({\bf 3}, {\bf 3})$, so that 
the model is anomaly free. 
(ii) In order to obtain the minus sign of cancellation, 
the family gauge boson masses are given by an inverted mass hierarchy 
$$
M^2(A_i^{\ j}) \equiv M_{ij}^2 = 
k \left( \frac{1}{m_{ei}^n} +  \frac{1}{m_{ej}^n} \right) + \cdots ,
\eqno(2.5)
$$
where ``$+\cdots$" denotes contributions from other scalars which
are negligibly small. 
(Here, although the number $n$ is $n=1$ in the original K-Y model
\cite{K-Y_PLB12}, we have denoted an extended case with $n \neq 1$ 
for convenience of later discussion.)  
Note that although only one scalar $\Phi$ gives charged lepton masses and
family gauge boson masses in the Sumino model \cite{Sumino_PLB09}, while, 
in the K-Y model \cite{K-Y_PLB12}, there are two scalars 
$\Psi$ and $\Phi$ which are $({\bf 3}, {\bf 3}')$ of U(3)$\times$U(3)$'$.
Only $\Phi$ can gives charged lepton masses as 
$m_{ei}\delta_i^{\ j} \propto \langle \Phi_i^{\ \alpha} \rangle  
\langle \bar{\Phi}_\alpha^{\ j} \rangle$.
On the other hand, only $\Psi$ contributes dominantly gauge boson masses, i.e. 
$M_{ij}^2 \propto  \langle \Psi_i^{\ \alpha} \rangle 
\langle \bar{\Psi}_\alpha^{\ j} \rangle$ with 
$\langle \Psi \rangle \langle\bar{\Phi} \rangle = k {\bf 1}$.
(For a case of $n \geq 2$, see later.) 
Therefore, the Sumino cancellation mechanism is satisfied only approximately.

In the present investigation, it is essential that the family gauge boson 
interactions are given by Eq.(1.1). 
The interaction (1.1) has been derived from the following scenario:
The family symmetry breaking is not caused by scalars ${\bf 3}$ and/or 
 ${\bf 6}$ of U(3), but it is caused by a scalar $({\bf 3}, {\bf 3}^*)$ of
U(3)$\times$U(3)$'$, which are broken at $\Lambda$ and
$\Lambda'$ ($\Lambda \ll\Lambda'$), respectively.
(In the original Sumino model, the scalar was $({\bf 3}, {\bf 3})$ of 
U(3)$\times$O(3).  
In the present investigation, the difference is not 
essential.) 
Therefore,  a direct gauge boson mixing
$A_i^{\ j} \leftrightarrow A_j^{\ i}$ ($i=1,2,3$) 
does not appear in this model. 
The U(3)$\times$U(3)$'$ is dominantly broken by a scalar $\Psi_i^\alpha$ 
which is $({\bf 3}, {\bf 3}^*)$ of U(3)$\times$U(3)$'$, i.e. by 
a vacuum expectation value (VEV)   
$\langle \Psi_i^\alpha \rangle = v_i \delta_i^\alpha$ 
as in the K-Y model \cite{K-Y_PLB12}.
In the limit of $\Lambda' \gg \Lambda$, we obtain the U(3) 
family current interaction (1.1). 
In the quark sector, since quark mass matrices $M_u$ 
and $M_d$ are, in general, not always diagonal on the diagonal basis 
of $M_e$, so that family number violations at tree level are caused   
only through the mixing matrices among up- and down-quarks, 
$U^u \neq {\bf 1}$ and  $U^d \neq {\bf 1}$.

On the other hand, 
the gauge boson masses $M_{ij}$ are also dominantly generated by VEV 
of scalar $\Psi_i^\alpha$ which is $({\bf 3}, {\bf 3}^*)$ of 
U(3)$\times$U(3)$'$, and whose VEV is given by 
$\langle \Psi_i^\alpha \rangle = \delta_i^\alpha v_i$.
Then, we obtain family gauge boson masses 
$$
M^2(A_i^{\ j}) \equiv M^2_{ij} = \frac{1}{2} g_A^2 (|v_i|^2 + |v_j|^2) + \cdots,
\eqno(2.6) 
$$
where ``$+\cdots$" denotes contributions from other scalars which
are negligibly small, 
so that the family gauge boson masses $M_{ij} \equiv M(A_i^{\ j})$ 
approximately satisfy relations
$$
2 M_{ij}^2 \simeq M_{ii}^2 + M_{jj}^2 .
\eqno(2.7)
$$
Here, the assumption $|\langle \Psi \rangle |^2 \gg
|\langle \Phi \rangle |^2$ is essential.
For example, in a case B$_1$ which is discussed later,
we consider that the largest component of $\langle \Phi \rangle$
is of an order of $10^2$ GeV, while the largest component of 
$\langle \Psi \rangle$ is of an order of $10^7$ GeV,

In the present paper, we investigate the following two Cases
A and B which satisfy the Sumino cancellation mechanism: 
Case A with an inverted mass hierarchy and Case B with 
a normal gauge boson mass hierarchy. 
In both cases, the gauge boson masses are given by Eq.(2.6), 
so that the gauge boson masses satisfy the relation (2.7). 
Since we still consider $m_{ei} \delta_i^{\ j} \propto 
\langle \Phi_i^{\ \alpha} \rangle \langle \bar{\Phi}_{\alpha}^{\ j} \rangle$, 
the difference between Case A and Case B is only in a  
relation of the VEV $\langle \Psi \rangle$
to the VEV $\langle \Psi \rangle$.

\vspace{2mm}

{\bf Case A}: The inverted gauge boson mass hierarchy 
(K-Y model like) 

Charged lepton masses are given by Eq.(2.5).
Here,  we also consider cases with $n \neq 1$ 
in addition to the case with $n=1$ in the original K-Y model. 
For example, for $n=2$ we suppose 
$\langle \Phi_i^{\ \alpha} \rangle \langle \bar{\Psi}_\alpha^{\ j} \rangle 
 \langle {\Phi}_j^{\ \beta} \rangle \propto \langle {E}_i^{\ \beta} \rangle$, 
where $\langle \bar{E}_\alpha^j \rangle=v_E {\rm diag}(1,1,1)$.

The gauge boson masses satisfy the relation (2.7), 
mass ratios can be expressed as follow:
$$
M_{33} : M_{32} : M_{22} : M_{31}: M_{21} : M_{11} 
= 1: \sqrt{\frac{a^2+1}{2} } : a: \sqrt{\frac{b^2+1}{2} } : 
\sqrt{\frac{b^2+a^2}{2} } : b ,
\eqno(2.8)
$$
where
$$
a \equiv \frac{M_{22}}{M_{33}} = \left(\frac{m_\tau}{m_\mu} \right)^{n/2} , 
\ \ \ \  
b \equiv \frac{M_{11}}{M_{33}} = \left( \frac{m_\tau}{m_e} \right)^{n/2} .
\eqno(2.9)
$$

Sumino cancellation condition $g_F^2/2 =(3/2) \zeta  e^2$ in the K-Y model
is rewritten as
$$
\left( \frac{g_F}{\sqrt2} \right)^2 \simeq \frac{1}{n} \, 
\frac{3}{2}\, \zeta \, e^2 ,
\eqno(2.10)
$$
because of $\log M_{ii}^2 = - n \log m_{ei} + {\rm const}$.
Here, the coupling constant $g_F$ is defined by
$$
{\cal H}_{fam}^{K-Y} = \frac{g_F}{\sqrt2} \sum_{f=u,d,\nu,e} (\bar{f}^i \gamma_\mu f_j)
(A_i^{\ j})^\mu .
\eqno(2.11)
$$
(For convenience of comparison with the Sumino model, the coupling 
constant $g_F$ in the original K-Y model \cite{K-Y_PLB12} has been
changed into $g_F/\sqrt2$.)
Note that, differently from the original Sumino model,  
the cancellation  in the K-Y model is satisfied 
only approximately.  
The factor $\zeta$ in Eq.(2.10) is a fine tuning factor which gives 
$K(\mu) \simeq 2/3$ almost independently of $\mu$, and it is 
numerically given by $\zeta=1.752$ in the case of $n=1$.
  
\vspace{2mm}

 {\bf Case B}: The normal gauge boson mass hierarchy 
(the original Sumino model type)

Gauge boson masses are given by
$$
 M_{ij}^2 = k (m_{ei}^n + m_{ej}^n) .
\eqno(2.12)
$$
Although in the original Sumino model \cite{Sumino_PLB09}, 
the scalar $\Phi$ gives the gauge boson masses $M_{ij}$ and 
the charged lepton masses $m_{ei}$, in the present investigation,  
we also consider other possibilities in addition to 
the case with $n=1$.
For example, a case with $n=2$ is realized by a VEV relation 
$\langle \Psi_i^\alpha \rangle \langle \bar{E}_\alpha^j \rangle =
\langle \Phi_i^\alpha \rangle  \langle \bar{\Phi}_\alpha^j \rangle$.
Then, the cancellation condition is given by 
$$
\left( \frac{g_F}{\sqrt2} \right)^2 =\frac{2}{n} e^2 =
\frac{4}{n} \left( \frac{g_w}{\sqrt2} \right)^2
\sin^2 \theta_w ,
\eqno(2.13)
$$
because of $\log M_{ii}^2 =  n \log m_{ei} + {\rm const}$.

From Eq.(2.12), the gauge boson mass ratios are expressed by 
$$
M_{11} : M_{12} : M_{22} : M_{13}: M_{23} : M_{33} 
= 1: \sqrt{\frac{a^2+1}{2} } : a: \sqrt{\frac{b^2+1}{2} } : 
\sqrt{\frac{b^2+a^2}{2} } : b ,
\eqno(2.14)
$$
where
$$
a \equiv \frac{M_{22}}{M_{11}} = \left( \frac{m_\mu}{m_e} \right)^{n/2}, 
\ \ \ \  
b \equiv \frac{M_{33}}{M_{11}} = \left( \frac{m_\tau}{m_e} \right)^{n/2} .
\eqno(2.15)
$$

In the original Sumino model, the currents with an unwelcome form
as shown in Eq.(2.4) appear inevitably. 
We want less contribution of the family gauge bosons to the 
$P^0$-$\bar{P}^0$ mixing.  
Therefore, in the present investigation in Case B, we slightly 
change the original Sumino model into a modified model where 
leptons $\ell_i= (\nu_i, e^-_i)$ are still assigned to 
$(\ell_L, \ell_R)=({\bf 3}, {\bf 3}^*)$, while quarks $q_i =(u_i, d_i)$ 
are assigned to $(q_L, q_R)=({\bf 3}, {\bf 3})$,
so that the quark sector is anomaly free. 
In Case B, the gauge boson interactions are given by
$$
{\cal H}_{fam}^{(B)} = \frac{g_F}{\sqrt{2}} \left[ \sum_{f= \nu, e} 
\left( \bar{f}_L^i \gamma_\mu f_{Lj} - \bar{f}_{Rj} \gamma_\mu f_R^i
\right) +  \sum_{f=u,d} (\bar{f}^i \gamma_\mu f_j) \right] (A_i^{\ j})^\mu ,
\eqno(2.16)
$$
instead of Eq.(2.4).  
However, the lepton currents with the unwelcome form still appear.
(We will provide additional heavy leptons in order to remove 
anomaly in the lepton sector.) 

\vspace{2mm}

Finally we would like to emphasize that we assume that the family symmetry U(3) is assumed for
all cases, so that the condition between $g_F$ and $e$ is unchanged
in Case A (and also Case B).
(For example, Eq.(2.3) is satisfied model-independently in Case B.)
However, since the relations between 
$\langle \Psi \rangle$ and $\langle \Phi \rangle$ (i.e. between $M_{ii}$ and $m_{ei}$) 
are model-dependent even the family symmetry U(3) is assumed in common, 
so that in Eqs.(2.5), (2.8) - (2.10) and (2.12) - (2.15),
the factor $n$  has appeared model-dependently. 

 
 \vspace{5mm}
 
\noindent{\large\bf 3 \ Quark family arrangements and $P^0$-$\bar{P}^0$ mixing}

Effective quark current-current interactions with $\Delta N_{fam} = 2$
are given by 
$$
H^{eff} = \frac{1}{2} g_F^2 \left[ \sum_i \frac{ (\lambda_i)^2 }{M_{ii}^2} 
+ 2 \sum_{i<j} \frac{\lambda_i \lambda_j }{M_{ij}^2} \right]
(\bar{q}_k \gamma_\mu q_l) (\bar{q}_k \gamma^\mu q_l ) 
\eqno(3.1)
$$
where
$$
\lambda_1 = U^{q*}_{1k} U^q_{1l} ,  \ \ \lambda_2 = U^{q*}_{2k} U^q_{2l} ,  \ \ 
\lambda_3 = U^{q*}_{3k} U^q_{3l} .  
\eqno(3.2)
$$
For example, in a case of $K^0$-$\bar{K}^0$ mixing, $\lambda_i$ are given by
$$
\lambda_1=U^{d*}_{11} U^d_{12},  \ \ \ \lambda_2= U^{d*}_{21} U^d_{22} 
\ \ \ \lambda_3= U^{d*}_{31} U^d_{32}.
\eqno(3.3)
$$
These $\lambda_i$ with $k\neq l$ satisfy a unitary triangle condition
$$
\lambda_1 +\lambda_2 + \lambda_3 = 0 .
\eqno(3.4)
$$
We define the effective coupling constant $G^{eff}$ in the current-current
interaction as 
$$
G^{eff} = \frac{1}{2} g_F^2 \left[ \frac{\lambda_1^2}{M_{11}^2} +
 \frac{\lambda_2^2}{M_{22}^2} +  \frac{\lambda_3^2}{M_{33}^2} +
 2\left( \frac{\lambda_1 \lambda_2 }{M_{12}^2} +
 \frac{\lambda_2 \lambda_3 }{M_{23}^2} + \frac{\lambda_3 \lambda_1 }{M_{31}^2}
 \right) \right] .
\eqno(3.5)
$$
Note that all family gauge bosons contribute to the $P^0$-$\bar{P}^0$ mixing
as seen in Eq.(3.1).

In order to demonstrate numerical results, we tentatively assume 
$U^u \simeq {\bf 1}$ and $U^d \simeq V_{CKM}$ ($V_{CKM}$ is the 
Cabibbo-Kobayashi-Maskawa (CKM) quark mixing matrix \cite{CKM}).
Alternative case  with $U^u \simeq  V_{CKM}^\dagger$ and 
$U^d \simeq {\bf 1}$ can give no contributions to $K^0$-$\bar{K}^0$, 
$B^0$-$\bar{B}^0$ and $B_s^0$-$\bar{B}_s^0$ mixings, so that it is 
good news for the present purpose.  
However, the case brings a more severe constraint on the gauge boson masses
from the observed value of $D^0$-$\bar{D}^0$ mixing.

The assumption $U^d \simeq V_{CKM}$ leads to values of $\lambda_i$, 
$$
\lambda_1 \simeq 0.220, \ \ \ \lambda_2 \simeq- 0.219, \ \ \ 
\lambda_3 \simeq -0.00035 .
\eqno(3. 6)
$$
Therefore, in the limit of $\lambda_3 \simeq 0$ and 
$\lambda_1\simeq - \lambda_2$, 
we obtain approximate relation
$$
G^{eff}_K \simeq  \frac{g_F^2}{2}  \frac{\lambda_1^2}{M_{11}^2} 
\ \ \ \left( {\rm or}\ G^{eff}_K
\simeq  \frac{g_F^2}{2}  \frac{\lambda_2^2}{M_{22}^2} \right). 
\eqno(3.7)
$$
Thus, the $K^0$-$\bar{K}^0$ mixing put a severe constraint on 
the lower bound of the family gauge boson mass $M_{11}$ for 
$M_{11} <M_{22}$ (or $M_{22}$ for $M_{22} < M_{11}$)]. 
When we use the observed value \cite{PDG12} $\Delta m_K^{obs} 
=(3.484\pm 0.006)\times 10^{-18}$ TeV and a tentative standard model
(SM) value \cite{Bigi} $\Delta m_K^{SM} \sim 2 \times 10^{-18}$ TeV, 
we obtain a lower limit of the value $M_{11}/(g_F/\sqrt2)$ [or 
$M_{22}/(g_F/\sqrt2)$]  $\sim 340$ TeV, where we have used 
a vacuum-insertion approximation (with no QCD correction)
$$
\Delta m_K^{fam} =\frac{1}{6} G_K^{eff} f_K^2 f_K (1+ 2 S_K) ,
\eqno(3.8)
$$
and $S_K = m_K^2/(m_s +m_d)^2$.
If we give the parameters $a$ and $b$ in Eq.(2.9) [or (2.15)], 
we can estimate $G^{eff}$ without approximation (3.7).  
In the next section, we will calculate constraints for 
$M_{ij}/(g_F/\sqrt2)$ directly from Eq.(3.5) and by using $V_{CKM}$ 
with $CP$ violation phase. 

Here, note that the CKM matrix $V_{CKM}$ is defined in the generation  
basis $u_i =(u,c,t)$ and $d_i =(d, s, b)$.  
Therefore, the notations $M_{ij}$ in Eqs.(3.1) are different 
from those defined by the diagonal bases of the charged lepton 
mass matrix $M_e$. 
In this paper, we investigate various assignments of 
$q_i=(q_1, q_2,q_3)$. 
As far as quark sector is concerned, the use of generation basis 
$d_i=(d, s, b)$ is convenient. 
Therefore, hereafter, for example, for Case B$_1$ with family number
$d_i =(b,s,d)$ (the case is defined in the next section), we denote 
$M_{11}$, $M_{12}$, $M_{22}$, $\cdots$ as $M_{bb}$, $M_{bs}$, 
$M_{ss}$,  $\cdots$, respectively, 
in order to distinguish those from $M_{ij}$ defined in the 
family numbers.  
(For convenience, we use down-quark names as the quark family  
numbers.)
The physics is highly dependent on the quark family assignments.
The details are discussed in the next section.

\vspace{3mm}

\noindent{\large\bf 4 \ Which quark-family assignment is favorable ?}

We find that $K^0$-$\bar{K}^0$ mixing puts the most severe constraints 
on the family gauge boson masses $M_{i}$ compared with other 
$P^0$-$\bar{P}^0$ mixings.  
As seen in Eqs.(3.6) and (3.7), because of $|\lambda_b|^2 \ll |\lambda_s|^2
\simeq |\lambda_d|^2$,  the observed  $K^0$-$\bar{K}^0$ mixing put a constraint 
on $M_{dd}$ or $M_{ss}$, but it does not put a constraint on $M_{bb}$.  
Therefore, for our purpose of visible family gauge bosons, we should 
regard the third generation quark $(t,b)$ as $(t,b)=(u_3, d_3)$ in 
Case A with the inverted gauge boson mass hierarchy,  and  
$(t,b)$ as $(t,b)=(u_1, d_1)$ in Case B with the normal gauge boson 
mass hierarchy. 
As a result, we have the following four candidates of the quark family assignments:
Case A$_1$:  $(d_1, d_2, d_3)=(d, s, b)$,  Case A$_2$:  $(d_1, d_2, d_3)=(s, d, b)$,  
Case B$_1$:  $(d_1, d_2, d_3)=(b, s, d)$ and Case B$_2$:  $(d_1, d_2, d_3)=(b, d, s)$.
Cases A$_1$ with $n=1$ and B$_1$ 
with $n=1$ correspond to the K-Y model and the Sumino model, 
respectively. 

In Table 1, we list gauge boson masses $M_{ij}$ estimated from 
$\Delta m_K^{fam} \sim 1.4 \times 10^{-8}$ TeV for these four cases 
with typical values of $n$, where $n$ is defined by Eq.(2.9) [or Eq.(2.15)].  
[Exactly speaking, since the value 
$\Delta m_K^{fam} \sim 1.4 \times 10^{-8}$ TeV
means those which we can take as large as possible, the values of 
$\tilde{M}_{ij}$ in Table 1 ($\tilde{M}_{ij}$ are defined  by Eq.(4.1) below) 
denote lower limits of $\tilde{M}_{ij}$.] 
In the evaluations of $\lambda_i$ we have taken not only of the magnitudes
of $V_{CKM}$ elements, but also the $CP$ phase \cite{PDG12} into 
consideration.   
In Table 1, for convenience, numerical values of masses are given by
$$
\tilde{M}_{ij}^2 \equiv  \frac{M_{ij}^2}{g_F^2/2} .
\eqno(4.1)
$$
As far as we treat four-Fermi current-current interactions,
the value $\tilde{M}_{ij}$ are practically useful rather 
than $M_{ij}$. 
Real mass values $M_{ij}$ are needed only when we discuss a direct 
observation of $A_i^j$.
In the numerical estimates of $\tilde{M}_{ij}$, note that 
the expression $M_{ij}$ given by Eq.(2.8) [and also Eq.(2.14)] have been 
described in the family numbers which defined by 
$(e_1, e_2, e_3)=(e^-, \mu^-, \tau^-)$, while the formula (3.5) with
Eq.(3.2) have been described by using the quark generation-number 
$(d_1, d_2, d_3)= (d, s, b)$.


\begin{table}
\caption{
Family gauge boson masses estimated from 
$\Delta m_K^{fam}\sim 1.4 \times 10^{-8}$ TeV. 
Here, we have used parameter values 
$a=(m_\tau/m_\mu)^{n/2}=(16.8167)^{n/2}$
and $b=(m_\tau/m_e)^{n/2} =(3477.15)^{n/2}$ for Case A, 
and $a= (m_\mu/m_e)^{n/2} = (206.768)^{n/2}$ and 
$b=(m_\tau/m_e)^{n/2} = (3477.15)^{n/2}$
for Case B.  
In this table, for convenience, numerical values of masses are given 
by $\tilde{M}_{ij} \equiv M_{ij}/(g_F/\sqrt2)$ in a unit of TeV. 
}

\vspace{2mm}

\begin{center}
\begin{tabular}{|c|ccccccccccc|} \hline
Case & \multicolumn{11}{|c|}{ Family gauge boson masses} \\ \hline
(A) & $M_{11}$ & $>$ & $M_{12}$ & $>$ & $M_{13}$ & $>$ & $M_{22}$ & 
$>$ & $M_{23}$ & $>$ & $M_{33}$ \\ 
Ratios &  $b$ &  &  $\sqrt{\frac{b^2+a^2}{2} }$ &  &  $\sqrt{\frac{b^2+1}{2} }$ &
  & $a$ & &  $\sqrt{\frac{a^2+1}{2} }$ & & 1\\ \hline 
(A$_1$) &  $\tilde{M}_{dd}$ & $>$ & $\tilde{M}_{ds}$ & $>$ & $\tilde{M}_{db}$ & 
$>$ & $\tilde{M}_{ss}$ & $>$ & $\tilde{M}_{sb}$ & $>$ & $\tilde{M}_{bb}$ \\ 
$n=1/2$ & 1209 & & 884.5 & & 862.5 & & 319.0 & & 251.5 & & 157.5   \\
$n=1$ &  5062 &  & 3588 & & 3580 & & 352.0 & & 256.2 & & 85.8 \\
$n=2$ &  73342 & & 51861 & & 51860 & & 354.7 & & 251.3 & & 21.1 \\ 
$n=3$ & $1.1\times 10^6$  & & $7.4\times 10^5$ &  & $7.4\times 10^5$ & & 
356.0 & & 251.8 & & 5.16 \\ \hline
(A$_2$) &  $\tilde{M}_{ss}$ & $>$ & $\tilde{M}_{sd}$ & $>$ & $\tilde{M}_{sb}$ & 
$>$ & $\tilde{M}_{dd}$ & $>$ & $\tilde{M}_{db}$ & $>$ & $\tilde{M}_{bb}$ \\ 
$n=1/2$ & 1205 & & 881.4 & & 859.5 & & 317.8 & & 250.7 & & 156.0 \\
$n=1$ & 5042 & & 3574 & & 3566 & & 350.7 & & 255.2 & & 85.5 \\
$n=2$ & 73035 & & 51644 & & 51644 & & 353.2 & & 250.2 & & 21.0 \\ 
$n=3$ & $1.2\times 10^6$ & &  $7.5\times 10^5$ & & $7.5\times 10^5$  & & 
354.5 & & 250.7 & & 5.14 \\ \hline
(B) & $M_{11}$ & $<$ & $M_{12}$ & $<$ & $M_{22}$ & $<$ & $M_{13}$ & 
$<$ & $M_{23}$ & $<$ & $M_{33}$ \\ 
Ratios  &  $1$ &  &  $\sqrt{\frac{a^2+1}{2} }$ & & $a$  &  &  $\sqrt{\frac{b^2+1}{2} }$
  & & $\sqrt{\frac{b^2+a^2}{2} }$ & & $b$ \\ \hline 
(B$_1$) &  $\tilde{M}_{bb}$ & $<$ & $\tilde{M}_{bs}$ & $<$ & $\tilde{M}_{ss}$ & 
$<$ & $\tilde{M}_{bd}$ & $<$ & $\tilde{M}_{sd}$ & $<$ & $\tilde{M}_{dd}$ \\ 
$n=1/2$ & 63.5 & & 176.0 & & 240.7 & & 347.5 & & 384.4 & & 487.4 \\
$n=1$ & 22.5 & & 229.8 & & 324.2 & & 940.2 & & 967.6 & & 1329 \\
$n=2$ & 1.77 & & 258.3 & & 365.3 & & 4344 & & 4352 & & 6144 \\ \hline
(B$_2$) &  $\tilde{M}_{bb}$ & $<$ & $\tilde{M}_{bd}$ & $<$ & $\tilde{M}_{dd}$ & 
$<$ & $\tilde{M}_{bs}$ & $<$ & $\tilde{M}_{ds}$ & $<$ & $\tilde{M}_{ss}$ \\ 
$n=1/2$ & 63.1 & & 174.9 & & 239.2 & & 345.4 & & 382.0 & & 484.3 \\
$n=1$ & 22.4 & & 228.7 & & 322.7 & & 935.8 & & 963.1 & & 1323 \\
$n=2$ & 1.76 & & 257.3 & & 363.9 & & 4327 & & 4334  & & 6119 \\ \hline
 \end{tabular}
 \end{center}
 \end{table}


As seen in Table 1, Case A$_1$ and Case A$_2$ lead to large values 
of $\tilde{M}_{ij}$, so that these cases are not interesting to us.  
Case A with $n \geq 3$ can have $\tilde{M}_{33}$ smaller than a few TeV, 
but the case gives $\tilde{M}_{11} \sim 10^6$ TeV.  
Phenomenology in Case A$_1$ with $n=1$ has already been 
investigated in Refs.\cite{K-Y_PLB12,YK_PRD13}.
Phenomenology for Case B with $d_i=(d,s,b)$ has investigated 
in Ref.\cite{K-S-Y_PLB11}. 
The results for visible effects of the family gauge bosons was negative. 

We consider that Case B with $n=2$  is phenomenologically  
most attractive, because the lightest family gauge boson $A_1^{\ 1}$ 
has mass of an order of a few TeV which is visible at the LHC 
(remember $M_{11} = (g_F/\sqrt2) \tilde{M}_{ii}$).
Besides, even the heaviest gauge boson has, at most, 
a mass of an order of $10^4$ TeV.

\vspace{3mm}

\noindent{\large\bf 5 \ Phenomenology of the family gauge bosons in Cases  
B$_1$ and B$_2$ }

In this section, let us investigate phenomenology of the family gauge bosons
in  Cases B$_1$ and B$_2$ with $n=2$. 
From a point of view of model-building, too, the case $n=2$ is not so 
unlikely, because we can consider a VEV relation 
$\langle \Psi \rangle_i^{\ \alpha} = \langle \Phi \rangle_i^{\ \beta} 
\langle \bar{E \rangle}_\beta^{\ j} \langle \Phi \rangle_j^{\ \alpha}$,
where $\langle \bar{E \rangle} = {\bf 1}$. 
In this case, from Eq.(2.13), the gauge coupling constant $g_F/\sqrt2$ 
is given by
$$
\left. \frac{g_F}{\sqrt2} \right|_{n=2} = e =0.30684 ,
\eqno(5.1)
$$
where, for convenience, we have used \cite{PDG12} 
$\alpha(m_\tau) =1/133.471$.

\vspace{2mm}

{\bf 5.1 \  Direct production of the lightest gauge boson $A_1^{\ 1}$ } 

From the value given in Table 1 and the value (5.1), the mass of gauge boson
$A_1^{\ 1}$ is 
$$
M_{11} \simeq  0.543\ {\rm TeV} \ \ (0.540\ {\rm TeV})  \ \ 
{\rm for \ Case \ B}_1 \ \ [{\rm Case\ B}_2] , 
\eqno(5.2)
$$
It should be noted that the gauge boson $A_1^{\ 1}$ can interact 
only with the third generation quarks $(t,b)$, although it does 
with the first generation leptons $(\nu_e, e)$ for leptons. 
Therefore, the gauge boson $A_1^{\ 1}$ will be produced by 
gluon fusion (Fig.1) as
$$
p+p \rightarrow A_1^1 + b + \bar{b} +X \rightarrow 
e^+ e^- + X ,
\eqno(5.3)
$$
at the LHC.
(In future, we will also observe $A_1^{\ 1}$ production in the ILC
as $e^+ + e^- \rightarrow A_1^1 $.)

\begin{figure}[h]
\begin{center}
\begin{picture}(200,150)(0,0)
 \includegraphics[height=.22\textheight]{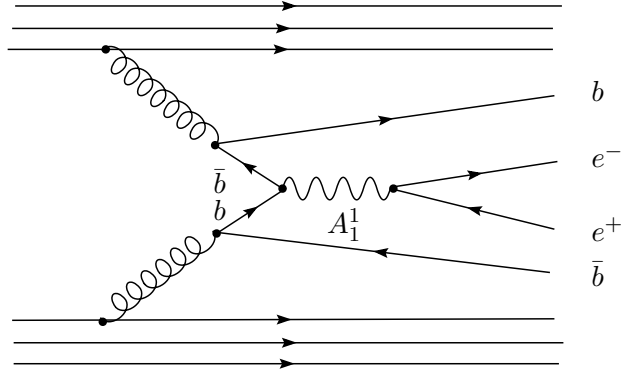}
 \put(-133,55){$b$}
  \put(-133,65){$\bar{b}$}
 \put(-90,50){$A_1^1$}
\put(10,100){$b$}
\put(10,75){$e^-$}
\put(10,47){$e^+$}
\put(10,30){$\bar{b}$}

\end{picture}  
  \caption{
$A_1^{\ 1}$ production at the LHC. 
}
\label{LHC}
\end{center}
\end{figure}


We have decay modes of $A_1^{\ 1}$ into $t+\bar{t}$, $b+\bar{b}$, 
$e^- + e^+$ and $\nu_e + \bar{\nu}_e$ with branching fractions 
as follows:
$$
\begin{array}{l}
Br( A_1^{\ 1} \rightarrow t \bar{t}) = 
Br( A_1^{\ 1} \rightarrow b \bar{b}) = \frac{6}{15} = 40 \% , \\  
Br( A_1^{\ 1} \rightarrow e^- e^+) = \frac{2}{15} =13.3 \% , \ \ 
Br( A_1^{\ 1} \rightarrow \nu_e \bar{\nu}_e) = \frac{1}{15}= 6.7 \% . 
\end{array}
\eqno(5.4)
$$ 
Note that the branching ratio $Br(A_1^{\ 1} \rightarrow \nu_e \bar{\nu}_e)
=1/15=6.7 \%$ is one in the case of Majorana neutrinos.
If neutrinos are Dirac neutrinos, the branching ratios is given 
$Br(A_1^{\ 1} \rightarrow \nu_e \bar{\nu}_e) =2/16=12.5 \%$. 
The large difference between both is due to the large leptonic 
branching ratio in the family gauge boson decays.
Therefore, in future, when data of the direct production of 
$A_1^{\ 1}$ are accumulated, we will be able to conclude 
whether neutrinos are Dirac or Majorana by observing whether 
$Br(A_1^{\ 1} \rightarrow \nu_e \bar{\nu}_e)$ is $6.7\%$ or $12.5\%$. 

The search for $A_1^{\ 1}$ production at the LHC is done by 
a similar way of the $Z'$ search (for a review, see, for example, 
\cite{review_Zsearch}). 
Although there has been an experimental report on $Z'$ search 
\cite{Z_serch_Tevatron}, the result cannot be applicable for
$A_1^{\ 1}$ search, because $A_1^{\ 1}$ cannot interact with 
the first generation quarks, so that the cross section is 
considerably small compared with $Z'$ production. 
The cross section of $A_1^{\ 1}$ in the original Sumino model 
has been discussed in Ref.\cite{K-S-Y_PLB11}, but
the case was a different family gauge boson
$A_1^{\ 1}$ which can interact with the first generation quarks.

Since the purpose of the present paper is to give an overview of 
the family gauge bosons with visible energy scale,  
estimate of the production rate $\sigma(pp\rightarrow A_1^1)$ will 
be given elsewhere.

If the real mass $M_{11}$ is smaller than 500 GeV, we may expect 
an observation at the ILC in future, too. 

\vspace{2mm}

{\bf 5.2 \  Contribution of family gauge bosons to the rare decay 
$K^+ \rightarrow \pi^+ \nu \bar{\nu} $}

Let us estimate contributions of family gauge bosons to the rare decay 
$K^+ \rightarrow \pi^+ \nu \bar{\nu} $, 
because only a finite value of the branching ratio has been reported 
 \cite{PDG12} at present:
$$
Br(K^+ \rightarrow \pi^+ \nu \bar{\nu})_{obs} =
(1.7 \pm 1.1) \times 10^{-10} .
\eqno(5.5)
$$ 
It is usually taken that this value is consistent with
the standard model prediction 
\cite{B-Kpinunu}
$$
Br(K^+ \rightarrow \pi^+ \nu \bar{\nu})_{SM} = (0.80 \pm 0.11)
\times 10^{-10} .
\eqno(5.6)
$$
We are interested in whether Case B is consistent or not
with the present experimental result (5.5).

In the present model, all family gauge bosons can, in principle, 
contribute to each rare decay mode. 
For example, in Cases B$_1$ and B$_2$, a transition $K \rightarrow \pi$
is mediated by the gauge bosons $A_s^{\ d} \equiv A_2^{\ 3}$ and  
$A_s^{\ d} \equiv A_3^{\ 2}$, respectively. 
However, as seen in Table 1, the mass of $M_{23}$ is of the order of 
$10^3$ TeV, so that the effect is invisible.  
Remember that family-number violating transitions are possible 
in the quark sector.
Since the effective mass value of $\tilde{M}_{11} \equiv \tilde{M}_{bb}$
is too small, the contribution of $A_1^{\ 1}$ is dominated compared with 
other gauge boson exchanges even considering the existence of 
the suppression factor $|U_{bd}^{d*} U_{bs}^d|$ (the value is $0.0155$ 
in the approximation $U^d \simeq V_{CKM}$). 
Then, the branching ratio due to the family gauge boson exchange $A_1^{\ 1}$
are estimated as follows:
$K^+ \rightarrow \pi^+ e^- \mu^+$ as follows:
$$
\frac{Br(K^+ \rightarrow \pi^+ \nu_e \bar{\nu}_e)_{fam}}{
Br(K^+ \rightarrow \pi^0 \mu^+ \nu_\mu)} = \eta_B \xi^2   
\frac{ f(m_{\pi^+}/m_K)}{\frac{1}{2}f(m_{\pi^0}/m_K)} (r_{11})^4 
\eqno(5.7)
$$
where
$$
(r_{ij})^2 =  \frac{ (g_F^2/2)/M_{ij}^2 }{(g_w^2/8)/M_W^2} 
= \frac{2 v_H^2}{\tilde{M}^2_{ij}} ,
\eqno(5.8)
$$
$v_H = 246\ {\rm GeV}$, and $f(x)$ is a phase space function 
$f(x)= 1-8 x^2 + 8 x^6 -x^8 -12 x^4 \log x^2$. 
(We have neglected the lepton masses.)
Here, the factor $\xi$ denotes mixing effects in quarks, and
in this case, $\xi$ is given by
$$
\xi = \frac{|V^*_{td} V_{ts} |}{|V_{us}|},
\eqno(5.9)
$$
where we have used the approximation $U^d \simeq V_{CKM}$. 
The factor $\frac{1}{2}$ in the denominator of Eq.(5.7) is due to
$\pi^0=(u\bar{u}-d\bar{d})/\sqrt{2}$. 
The factor $\eta$ denotes difference of effective current-current
interactions: When we denote the currents for weak interactions
$(\bar{\nu}\gamma_\mu (1-\gamma_5) e)$ as $(V-A)$ symbolically,
the factor $\eta_B$ for a final state of $\nu\bar{\nu}$ is given 
by $\eta_B^{\nu \nu}=[(|V|^2+|A|^2)/4]/(|V|^2+|A|^2) =1/4$ 
because only the left-handed neutrino $\nu_L$ can contribute 
as seen in Eq.(2.16).   
[In contrast to the case $\nu \bar{\nu}$, for a final state of $e^+ e^-$, 
it is given by $\eta_B^{ee}= 1/2$.] 

We obtain numerical results
$$
Br(K^+ \rightarrow \pi^+\nu_e \bar{\nu}_e)_{fam} = 1.1 \times 10^{-10} 
\ (0.91 \times 10^{-10}) \  \   {\rm for} 
\ \tilde{M}_{11} = 1.8\ (1.9) \ {\rm TeV} ,
\eqno(5.10)
$$
 form Eq.(5.7). 
This value is just favorable to the difference between the observed one (5.5)
and the SM one (5.6), i.e. $(1.7 -0.8)\times 10^{-10}$.  
(However, it should be noted that result (5.10) is only an approximate one, 
because we have neglected interference with the final state mode 
from the standard model. 
The numerical result should be taken rigidly.)

For rare $B$ and $K$ decays, we can estimate their branching ratios by 
a similar way to Eq.(5.7).
We investigate only the decay modes via the family gauge boson 
$A_2^{\ 1}$, because other gauge bosons are considerably 
heavy, so that such gauge boson effects are obviously invisible.
Note that since the family number in the quark sector is assigned 
unconventionally, for example, the gauge boson $A_2^{\ 1}$ causes 
the decay $B\rightarrow K + e^+ + \mu^-$ 
with mixing factor $\xi=|U_{bb} U_{ss}|/|V_{us}|$ for Case B$_1$ 
with $A_1^{\ 2} \equiv A_b^{\ s}$, and 
$B\rightarrow \pi + e^+ + \mu^-$ 
with mixing factor $\xi=|U_{bb} U_{dd}|/|V_{us}|$ for Case B$_2$
with $A_1^{\ 2} \equiv A_b^{\ d}$. 
Differently from the decay $K \rightarrow \pi \nu_e \bar{\nu}_e$, 
the lightest gauge boson $A_1^{\ 1}$ cannot contribute. 
Since rare $B$ and $K$ decays via the lightest family gauge boson 
$A_1^{\ 1}$ yields final states $e^+ e^-$ and $\nu_e \bar{\nu}_e$, 
such decay modes are confused with decay modes via photon and 
$Z$ boson. 
The lightest gauge boson $A_i^{\ j}$ with $i\neq j$ is $A_2^{\ 1}$.
The branching ratios of decay modes via $A_2^{\ 1}$ are, for example, 
as follows:
$$
\begin{array}{lll}
{\rm Case\ B}_1 : \ \ & 
Br(B^+\rightarrow K^+ \mu^- e^+)\simeq 2.1 \times 10^{-11}, &
Br(B^0\rightarrow K^0 \mu^- e^+)\simeq 2.1 \times 10^{-11}, \\
{\rm Case\ B}_2 : \ \ & 
Br(B^+\rightarrow \pi^+ \mu^- e^+)\simeq 2.1 \times 10^{-11}, &
Br(B^0\rightarrow \pi^0 \mu^- e^+)\simeq 1.0 \times 10^{-11},  \\
\end{array}
\eqno(5.11)
$$ 
where we have assumed $U^d \simeq V_{CKM}$.
These results are invisible for a time, because the present 
experimental lower limits \cite{PDG12} are 
$Br(B^+\rightarrow K^+ \mu^- e^+) < 9.1 \times 10^{-8}$ and 
$Br(B^+\rightarrow \pi^+ \mu^\mp e^\pm) < 1.7 \times 10^{-7}$.
The family gauge boson $A_2^{\ 1}$ can also contribute to rare 
$K$ decays.
However, the predicted branching ratios are of orders of
$10^{-15} - 10^{-17}$ because small values of quark mixing 
factors, so that the effects invisible.  

\vspace{2mm}

{\bf 5.3 \  $\mu^- + N(A,Z) \rightarrow e^-+ N(A,Z)$ } 

So far, 
phenomenological merits of Cases B$_1$ and B$_2$ has been almost equal.  
In this subsection, we would like to emphasize that 
$\mu^- N \rightarrow e^- N$ is visible in Case B$_2$, while
it is invisible in Case B$_1$. 
 
Most sensitive test in the near future for  Cases B$_1$ and B$_2$ 
is to observe the so-called $\mu$-$e$ conversion. 
(For a review of the $\mu$-$e$ conversion and  more detailed calculations,
for example, see Ref.\cite{Kuno-Okada_RMP01} and 
Ref.\cite{Kitano_PRD02}, respectively.) 
The present experimental limit is, for instance, for $Au$, \cite{SINDRUM06} 
$$
R(Au) \equiv \frac{\sigma(\mu^- +Au \rightarrow e^- + Au)}{
\sigma(\mu\ {\rm capture}) } < 7\times 10^{-13} .
\eqno(5.12)
$$

The reaction $\mu^- N \rightarrow e^- N$ is caused by an exchange of 
the family gauge boson $A_2^{\ 1}$.
It means the exchange of $ A_2^{\ 1}\equiv A_s^{\ b}$ 
[$ A_2^{\ 1}\equiv A_d^{\ b}$] in Case B$_1$  [Case B$_2$].
At present, we do not know values of $|U_{ij}^q|$ ($q=u,d$).
Therefore, it is not practical, at this stage, to estimate a $\mu$-$e$ 
conversion rate strictly. 
Instead, we roughly estimate a $\mu$-$e$ conversion rate
in the quark level as follows: 
$$
R_q \equiv \frac{ \sigma(\mu^- + q \rightarrow e^- + q)}{
\sigma( \mu^- + u \rightarrow \nu_\mu + d) }
\simeq \left( \xi \frac{g_F^2/2}{\tilde{M}_{12}^2} 
\frac{M_W^2}{g_w^2/8} \right)^2 = 
\left( \xi (r_{12})^2 \right)^2,
\eqno(5.13)
$$
where $q=u, d$, and $(r_{12}^2)$ is defined by Eq.(5.8). 
(Although the estimated value $R_q$ has different physical meaning  
from the value $R(Au)$, we consider that the order of the value $R_q$
can provide one with useful information.)
In Eq.(5.13), $\xi$ is a quark mixing factor similar to Eq.(5.9), 
and the value of $\xi$ is given by 
$\xi =|U^{d*}_{sd} U^d_{bd}|/|V_{ud}|=2.00 \times 10^{-3}$ 
[$\xi =|U^{d*}_{dd} U^d_{bd}|/|V_{ud}|)=0.867 \times 10^{-2}$]  
 in Case B$_1$  [in Case B$_2$] under the approximation 
$U^d \simeq V_{CKM}$. 
In this approximation, we may regard 
the ratios $R_q$ as $R_u \ll R_d$, so that we can neglect contribution 
to nucleon from $R_u$ compared with that from $R_d$. 
Then, we can roughly estimate values of $R_q$ 
$$
R_q \simeq 
R_d \sim 1.32 \times 10^{-17} \ ( 2.52 \times 10^{-16}) \ \ 
{\rm for\ Case\ B}_1 \ ({\rm Case\ B}_2),
\eqno(5.14)
$$
where we have used $\tilde{M}_{12}=260$ TeV from Table 1.

In the near future, the COMET experiment \cite{Comet} 
will reach a single-event sensitivity of $2.6 \times 10^{-17}$.
Therefore, the value $R_q \sim 10^{-16}$ in Case B$_2$ 
become within reach of our observation, but the value 
$R_q \sim 1.32 \times 10^{-17}$ in Case B$_1$ is critical for 
its observation. 

Since the decay $\mu^- \rightarrow e^- +\gamma$ is highly 
suppressed in the present scenario, if we observe 
$\mu^- N \rightarrow e^- N$ without observation of 
$\mu^- \rightarrow e^- +\gamma$, 
then it will strongly support our family gauge boson scenario.
(The decay $\mu^- \rightarrow e^- +\gamma$ can occur through a 
quark-loop diagram.
However, such a diagram is highly suppressed.) 
%

\vspace{2mm}

{\bf 5.4 \  Deviations from the $e$-$\mu$-$\tau$ universality}

Previously, we pointed out \cite{YK_PRD13} a possibility of 
a deviation from the $e$-$\mu$ universality in tau decays 
$\tau \rightarrow \mu \nu \bar{\nu}/e \nu \bar{\nu}$ 
by assuming $\tilde{M}_{23} \ll \tilde{M}_{31}$. 
However, in the present model, we cannot observe such a deviation 
because the mass spectrum in the present model gives 
$\tilde{M}_{23} \simeq \tilde{M}_{31}$, and besides, we have
a large value $\tilde{M}_{23} \sim 10^3$ TeV in Case B.  

On the other hand, we have a possibility of sizable deviations 
from the $e$-$\mu$-$\tau$ universality in the $\Upsilon$ decays
$\Upsilon \rightarrow \tau^+ \tau^- / \mu^+ \mu^- /e^+ e^-$, 
because the value of $\tilde{M}_{11} \equiv \tilde{M}_{bb}$ is considerably 
small in Case B.
We have matrix elements for the decays 
$\Upsilon \rightarrow \tau^+ \tau^- / \mu^+ \mu^- /e^+ e^-$, 
as follows: ${\cal M}_{\tau\tau} = {\cal M}_{\mu\mu} \equiv {\cal M}_{SM}$
and 
${\cal M}_{ee} = {\cal M}_{SM} + {\cal M}_{fam} = {\cal M}_{SM} (1 -\varepsilon)$, 
where
$$
\varepsilon \simeq \frac{g_F^2/2}{(e/3)^2} \frac{M_\Upsilon^2}{M_{11}^2 -M_\Upsilon^2}
\simeq \frac{9}{e^2} \frac{ M_\Upsilon^2}{\tilde{M}_{11}^2} =2.64 \times 10^{-3} .
\eqno(5.15)
$$
Therefore, we can expect a deviation 
$$
1 - \frac{Br(\Upsilon \rightarrow e^+ e^-)}{Br(\Upsilon \rightarrow \mu^+\mu^-)} 
\simeq 2 \varepsilon = 0.0053. 
\eqno(5.16)
$$ 
At present, we have not observed such a deviation \cite{PDG12}.
However, the value (5.16) will become visible in future experiments.

\vspace{3mm}

\noindent{\large\bf 6 \ Concluding remarks}

We have investigated possibility of visible family gauge boson effects
for six family assignments in the quark sector  
$(d_1, d_2, d_3)=(d, s, b)$,  $(d_1, d_2, d_3)=(b, d, b)$, and so on,
under the Sumino cancellation condition. 
In the Sumino model, the family number is defined by the diagonal
basis of the charged lepton mass matrix $M_e= {\rm diag} (m_e, m_\mu, m_\tau)$.
The $P^0$-$\bar{P}^0$ mixings ($P=K, D, B, B_s$) are caused only through 
quark mixings $U^u \neq {\bf 1}$ and $U^d \neq {\bf 1}$. 
We have found that the most interesting case is Case B$_2$,  
$(d_1, d_2, d_3)=(b, d, s)$. 
In Case B$_2$, a direct production of $A_1^{\ 1}$ at the LHC, 
$\mu$-$e$ conversion $\mu^- N\rightarrow e^- N$, and  a deviation from
$e$-$\mu$-$\tau$ universality in the $\Upsilon$ decay will be observed 
in future experiments.
Also, Case B$_1$, $(d_1, d_2, d_3)=(b, s, d)$, is attractive, 
although the case is somewhat hard to observe in 
$\mu^- N\rightarrow e^- N$ compared with Case B$_2$.

In Case B, the leptons take a Sumino-like structure (so that 
Sumino's cancellation mechanism is satisfied), while quarks 
takes a twisted family-number assignment. 
At present, there is no theoretical ground for such family-number 
assignments.  
In order to make the twisted family-number assignment 
$(d_1, d_2, d_3)=(b,d,s)$ more reliable, 
we, at least, have to build a unified mass matrix model of quarks and 
leptons under such the twisted family-number assignment. 
It is a task in future. 

We hope that many physicists turn their attention 
to a possibility of visible family gauge bosons 
and of a twisted family-number assignment versus generation-numbers.

 \vspace{3mm}

{\Large\bf Acknowledgments} 

The author thanks T.~Yamashita  
for valuable and helpful conversations. 
He also thanks H.~Yokoya for helpful comments on 
the direct production of $A_i^{\ j}$ at the LHC,  
Y.~Kuno and H.~Sakamoto for useful comments on  
experimental status of $\mu$-$e$ conversion, and 
M.~Tanimoto for valuable information on the recent estimates of 
$K^0$-$\bar{K}^0$ mixing. 


%

%

\end{document}